\documentclass[12pt,preprint]{aastex}
\shorttitle{Balmer-Dominated Shocks}
\shortauthors{Heng \& McCray}
\begin{document}

\title{BALMER-DOMINATED SHOCKS REVISITED}

\author{Kevin Heng\altaffilmark{1} \& Richard McCray\altaffilmark{1}}

\altaffiltext{1}{JILA, University of Colorado, 440 UCB, Boulder, CO 80309-0440; hengk@colorado.edu, dick@jila.colorado.edu.}

\begin{abstract}
We present a new formalism to describe the ratios and profiles of emission lines from hydrogen in Balmer-dominated shocks.  We use this model to interpret the measured widths and ratios of broad and narrow H$\alpha$, H$\beta$ and Ly$\alpha$ emission lines in supernova remnants (SNRs).  Our model results agree fairly well with those obtained previously by Chevalier, Kirshner \& Raymond (1980) and are consistent with observations of several SNRs.  The same model fails to account for the ratio of broad to narrow line emission from the reverse shock in SNR 1987A as observed by Heng et al. (2006).  We suggest that this discrepancy between theory and observation results from a faulty assumption that Balmer-dominated shocks can be treated as sharp discontinuities.  If the spatial structure of the shock transition zone is taken into account, the predicted ratios of broad to narrow line emission in most SNRs will change by modest factors, but the ratio in SNR 1987A will increase substantially.  Significantly greater shock velocities will be required to account for the observed full widths at half-maximum of the broad emission lines in most SNRs.
\end{abstract}

\keywords{methods: shock waves --- supernova remnants}

\section{INTRODUCTION}
\label{sect:intro}
When a fast astrophysical shock enters neutral interstellar gas, we see an optical spectrum that is dominated by H$\alpha$ and other Balmer lines and is nearly devoid of the forbidden lines typically seen in emission line nebulae.  The Balmer emission is caused by impact excitation of neutral hydrogen atoms by fast ions and electrons in the shocked gas.  The emission lines in such ``Balmer-dominated'' or ``non-radiative'' shocks have two components.  The first is a narrow component with a line width characteristic of the cold interstellar gas, which results from direct excitation of the neutral hydrogen atoms.  The second is a broad component with a line width characteristic of the thermal velocity broadening of the shocked protons.  The broad component results from charge transfer reactions, which pass electrons from the nearly stationary hydrogen atoms to the shocked protons.  The resulting fast hydrogen atoms may be created in excited states or may be excited by subsequent collisions with other fast ions and electrons.  The resulting ratio of the broad to narrow components and the profiles of the broad Balmer lines depend on the equilibration of electron and ion temperatures in the shocked gas, on resonance trapping of the Lyman lines, and on the velocity and the inclination of the shock surface.  

The theory to interpret the emission spectra of such Balmer-dominated shocks was developed originally by Chevalier \& Raymond (1978, hereafter CR78) and Chevalier, Kirshner \& Raymond (1980, hereafter CKR80) and augmented by Smith et al. (1991, hereafter S91) and Ghavamian et al. (2001, hereafter G01).  It has been used to interpret the emission spectra of several supernova remnants (SNRs), including: SN 1006 (CKR80; Kirshner, Winkler \& Chevalier 1987, hereafter KWC87; S91; Winkler \& Long 1997; Ghavamian et al. 2002, hereafter G02); Kepler (Fesen et al. 1989, hereafter F89; Blair, Long \& Vancura 1991, hereafter BLV91; Sollerman et al. 2003; Sankrit et al. 2005); Tycho/SN1572 (KWC87; S91; Ghavamian et al. 2000; G01), RCW 86 (Long \& Blair 1990, Smith 1997; G01; S03), portions of the Cygnus Loop (Raymond et al. 1983; Fesen \& Itoh 1985; Hester, Raymond \& Blair 1994; G01), and four remnants in the Large Magellanic Cloud (Tuohy et al. 1982; S91; Smith et al. 1994).  Generally, the theoretical models fit the observations fairly well, but significant discrepancies (by factors $\sim 2$) between the theoretical and observed broad-to-narrow line ratios persist in some cases, as we shall discuss.

Supernova 1987A has provided a dramatic new example of a Balmer-dominated shock.  In contrast to the SNRs cited above, broad Balmer emission from SNR 1987A comes from hydrogen atoms in the rapidly expanding ($\sim 12,000$ km~s$^{-1}$) supernova debris rather than the stationary interstellar gas.  In this case, the highest velocity component of the emission lines, called ``surface emission,'' results from the excitation of freely-streaming hydrogen atoms that cross the reverse shock. Observations of the profiles of H$\alpha$ and Ly$\alpha$ emission from the reverse shock with the Space Telescope Imaging Spectrograph (STIS) have enabled us to measure the rapidly evolving flux of hydrogen atoms across the reverse shock and to reconstruct partially the three-dimensional geometry of the shock surface (Michael et al. 1998a, b, 2003, with the last hereafter M03; Heng et al. 2006, hereafter H06). In addition to the surface emission, H06 identified a lower velocity component of the H$\alpha$ and Ly$\alpha$ emission lines, which they called ``interior emission.'' Evidently, this interior emission results from charge transfer of electrons from the freely-streaming hydrogen atoms to protons in the shocked gas and is analogous to the broad component seen in the Balmer lines of other SNRs.  We also note that Balmer-dominated bow-shocks have been observed in four pulsar nebulae (Bucciantini 2002).

Here, we revisit the theory of Balmer-dominated shocks.  We hope to resolve the discrepancies between observations and theory, to understand better how to interpret the line profiles, and to extend the existing theoretical models to the very high shock velocities seen in SNR 1987A.  In \S\ref{sect:assumptions}, we state the assumptions of our model and mention their limitations.  In \S\ref{sect:rates}, we discuss the cross sections used and present the resulting velocity distribution functions and rate coefficients.  In \S\ref{sect:tree}, we present a new formalism for deriving the intensities and the profiles of broad and narrow emission lines.  In \S\ref{sect:results}, we compare our results to previous models and to existing data on Balmer-dominated remnants and SNR 1987A.  Finally, in \S\ref{sect:discussion}, we discuss the limitations of the model and future development that is needed.

\section{MODEL AND ASSUMPTIONS}
\label{sect:assumptions}

The basic model we consider is a plane-parallel shock with velocity $v_s$ that strikes a stationary gas of cold neutral hydrogen and helium, with the latter having fractional abundance $\chi_{He} = n_{He}/n_H$.  (Other elements that may be present with typical cosmic abundances will have no significant effects.)  We assume that the shocked gas is fully ionized and that the ions and electrons comprise a fluid having a velocity $v_s/4$ relative to the shock frame.  We assume that the ions and electrons have Maxwellian velocity distributions.

There is an uncertainty regarding the equilibration of post-shock ion temperatures.  Electrons and protons are heated to temperatures having a minimum ratio equal to that of their masses (i.e., $T_e/T_p \sim 2000$).  The Coulomb equilibration time often exceeds the age of the remnant, in which case the temperatures will remain unequal throughout the transition zone where the neutral atoms become ionized.  However, plasma waves and magneto-hydrodynamical turbulence at the shock front may transfer energy rapidly from protons to electrons.  We consider two models.  In the first, ``Model F'', the electron and ion temperatures are fully equilibrated.  In the second, ``Model N'', two-stream plasma instabilities produced by protons reflected upstream result in $T_e \sim 0.25 T_p$ (Cargill \& Papapoulos 1988; see also references in G02).  Following G02, we define an equilibration parameter, $f_{eq}$, such that 
\begin{eqnarray}
T_p = \frac{3 m_p v^2_s}{16 k} \left[ \mu f_{eq} + 1 - f_{eq}\right],\nonumber\\
T_e = \frac{3 m_p v^2_s}{16 k} \left[ \mu f_{eq} + \frac{m_e}{m_p}\left(1 - f_{eq}\right)\right],
\end{eqnarray}
where $\mu \approx (1 + 4\chi_{He})/(2 + 3\chi_{He})$ is the mean molecular weight.  For example, $\mu \approx 0.61$ for Galactic abundances ($\chi_{He} = 0.1$).  Consequently, $f_{eq}=1$ for Model F and $f_{eq} \approx 0.35$ for Model N.  We further assume that equilibration between the protons and the alpha particles occurs so rapidly that their post-shock temperatures are equal, namely $T_p = T_\alpha$.

The most questionable assumption of this model is that the hydrogen atoms enter a shocked ion plasma having uniform density, velocity, and temperature given by the adiabatic jump conditions.  In reality, the hydrogen atoms are ionized and excited in a transition zone where the ions are produced, decelerated, and heated.  No model, including the present one, of emission from Balmer-dominated shocks has included a proper description of this transition zone.  In \S\ref{sect:discussion}, we discuss the consequences of this approximation further, deferring a more detailed treatment to a future paper.

\section{REACTION RATES AND CROSS-SECTIONS}

\label{sect:rates}

\subsection{FREELY-STREAMING ATOMS}
\label{subsect:maxdirac}

Consider a beam of freely-streaming, neutral hydrogen atoms in the ground state that crosses the shock and encounters a thermal plasma (with relative velocity $v_{sh} = 3v_s/4$).  Such an atom has three possibilities open to it --- electron or ion impact excitation (denoted ``$E$''), impact ionization (denoted ``$I$''), and charge transfer (denoted ``$T$'').  
Unless specified, when the label ``$T$'' is used it refers to charge transfer to both the ground and the excited states, while ``$T^*$'' denotes only the latter.  

The rate that a hydrogen atom will have a reaction $X$ (where $X$ stands for $E$, $I$, or $T$) with a particle of type $s$ (in units of s$^{-1}$) is given by
\begin{equation}
R(\vec{v}_{sh};\sigma_X) = n_a \int \int f_a(\vec{v}_a) ~f_b(\vec{v}_b) ~\sigma_X(\vert \vec{v}_a-\vec{v}_b \vert) ~\vert \vec{v}_a-\vec{v}_b \vert ~d^3v_a ~d^3v_b.
\end{equation}

In the case at hand, the ions (with the subscript $s$ denoting protons, electrons or alpha particles) are assumed to have mass $m_s$, number density $n_s$, temperature $T_s$, and a Maxwellian velocity distribution function,
\begin{equation}
f_M(\vec{v}) = f_{M,0} ~\exp\left(-\frac{m_s v^2}{2kT_s}\right),
\end{equation}
where $f_{M,0}=(m_s / 2 \pi kT_s)^{3/2}$. 
The beam of hydrogen atoms has the distribution function,
\begin{equation}
f_0(\vec{v}_H,\vec{v}_{sh}) = \delta(\vec{v}_H - \vec{v}_{sh}).
\end{equation}
The rate of the reaction $X$ between them is then
\begin{eqnarray}
R_{X_0}(\vec{v}_{sh};\sigma_X) = n_s f_{M,0} \int \int \exp\left(-\frac{m_s v^2}{2 k T_s}\right) ~\delta(\vec{v}_H - \vec{v}_{sh}) ~\sigma_X(\vert \vec{v}_H - \vec{v} \vert) ~\vert \vec{v}_H - \vec{v} \vert ~d^3v_H ~d^3v \nonumber\\
= 2 \pi n_s f_{M,0} \int_{-\infty}^{\infty} \int_0^{\infty} ~\exp\left[-\frac{m_s(v_r^2+v_z^2)}{2 k T_s}\right] ~\Delta v ~\sigma_X(\Delta v) ~v_r ~dv_r ~dv_z,
\label{eq:rate1}
\end{eqnarray}
in cylindrical coordinates, where $d^3v = 2\pi v_r ~dv_r ~dv_z$ and $\Delta v = \sqrt{v_r^2 + (v_z - v_{sh})^2}$.  The reaction rate coefficient is simply $\tilde{R}_{X_0}=R_{X_0}/n_s$.  We further define the fractional abundance, $\chi_s \equiv n_s/n_H$, such that $R_{X_0}/n_H = \chi_s \tilde{R}_{X_0}$.

A fraction of the incoming beam of hydrogen atoms will undergo charge transfer reactions with protons in the shocked plasma, producing hydrogen atoms having a distribution function given by 
\begin{equation}
f_1(\vec{v},\vec{v}_{sh}) = S_1({\vec{v}_{sh}}) ~f_M(\vec{v})~\vert\vec{v}-\vec{v}_{sh}\vert~\sigma_{T}(\vert\vec{v}-\vec{v}_{sh}\vert),
\end{equation}
where the charge transfer cross section, $\sigma_T$, includes reactions to all excited states as well as to the ground state.  The value of the normalization constant, $S_1(\vec{v}_{sh})$, is fixed by the condition $\int f_1 ~d^3v=1$.  If we define
\begin{equation}
\tilde{R}_{T_0}(\vec{v}_{sh};\sigma_T) = \int f_M(\vec{v}) ~\vert\vec{v}-\vec{v}_{sh}\vert ~\sigma_{T}(\vert\vec{v}-\vec{v}_{sh}\vert) ~d^3v,
\end{equation}
then it follows that $S_1 = \tilde{R}_{T_0}^{-1}$. 

The function $f_1(\vec{v},\vec{v}_{sh})$ describes the velocity distribution function of hydrogen atoms resulting from charge transfer reactions of the original hydrogen beam with protons in the ionized gas.  The subscript ``1'' denotes that the atoms are the result of one charge transfer reaction.  The distribution $f_1$ is not Maxwellian because it includes factors of the relative velocity between the beam and the ions and the charge transfer cross section.  For example, it has the value $f_1(\vec{v}_{sh},\vec{v}_{sh}) = 0$ at the velocity of the initial beam.
 
\subsection{SUBSEQUENT REACTIONS}
\label{subsect:subsequent}

These new hydrogen atoms can in turn be excited, ionized, or undergo another charge transfer reaction.  If they are excited, they will emit one or more photons and remain in the same distribution function.  

The velocity distribution function of hydrogen atoms that have undergone $i$ charge transfer reactions is given by  
\begin{equation}
f_i(\vec{v},\vec{v}_{sh}) = S_i(\vec{v}_{sh}) ~f_M(\vec{v}) \int ~f_{i-1}(\vec{v}^\prime,\vec{v}_{sh}) ~\vert\vec{v}-\vec{v}^\prime\vert ~\sigma_T(\vert\vec{v}-\vec{v}^\prime\vert) ~d^3v^\prime,
\end{equation}
where $S_i= \tilde{R}_{T_{i-1}}^{-1}$. 
The corresponding rate coefficients for reactions $X$ of such atoms are
\begin{equation}
\tilde{R}_{X_i}(\vec{v}_{sh};\sigma_X) = \int \int  f_M(\vec{v}) ~f_i(\vec{v}_H,\vec{v}_{sh}) ~\vert \vec{v}_H - \vec{v} \vert ~\sigma_{X}(\vert \vec{v}_H - \vec{v} \vert) ~d^3v_H ~d^3v.
\label{eq:rate2}
\end{equation}

\subsection{CROSS SECTIONS}
\label{subsect:cross}

To calculate the reaction rate coefficients for excitation, ionization, and charge transfer (by interactions with protons, electrons, and alpha particles), we use analytical fitting functions to the cross sections provided by Janev \& Smith (1993).

We must distinguish between the net charge transfer cross section, $\sigma_T$, and the cross sections, $\sigma_{T^*}$, for charge transfer reactions which leave the hydrogen atom in excited states (and consequently produce photons).  For the latter, we use the cross sections of Barnett (1990), who tabulates results for $n=2s$ and $n=2p$. To obtain cross sections for $n > 2$, we use the scaling relation (Janev \& Smith 1993; C.D. Lin 2006, private communication),
\begin{equation}
\sigma_{T^*,n} = \left(\frac{2}{n}\right)^3 ~\sigma_{T^*,2}.
\end{equation}
We include all excitation and charge transfer reactions up to and including $n=6$.

We also consider charge transfer reactions between alpha particles and hydrogen atoms, which produce singly-charged helium ions.  We include reactions to He$^+$ in the states $n=1s, 2s, 2p, 3s, 3p, 3d$ and $4p$, using the data of Barnett (1990). We do not follow the evolution of these ions; instead, we regard the reactions producing them as equivalent to impact ionization of hydrogen and include the corresponding cross sections as a contribution to the net impact ionization cross section.  

\begin{figure}
\begin{center}
\includegraphics[width=6in]{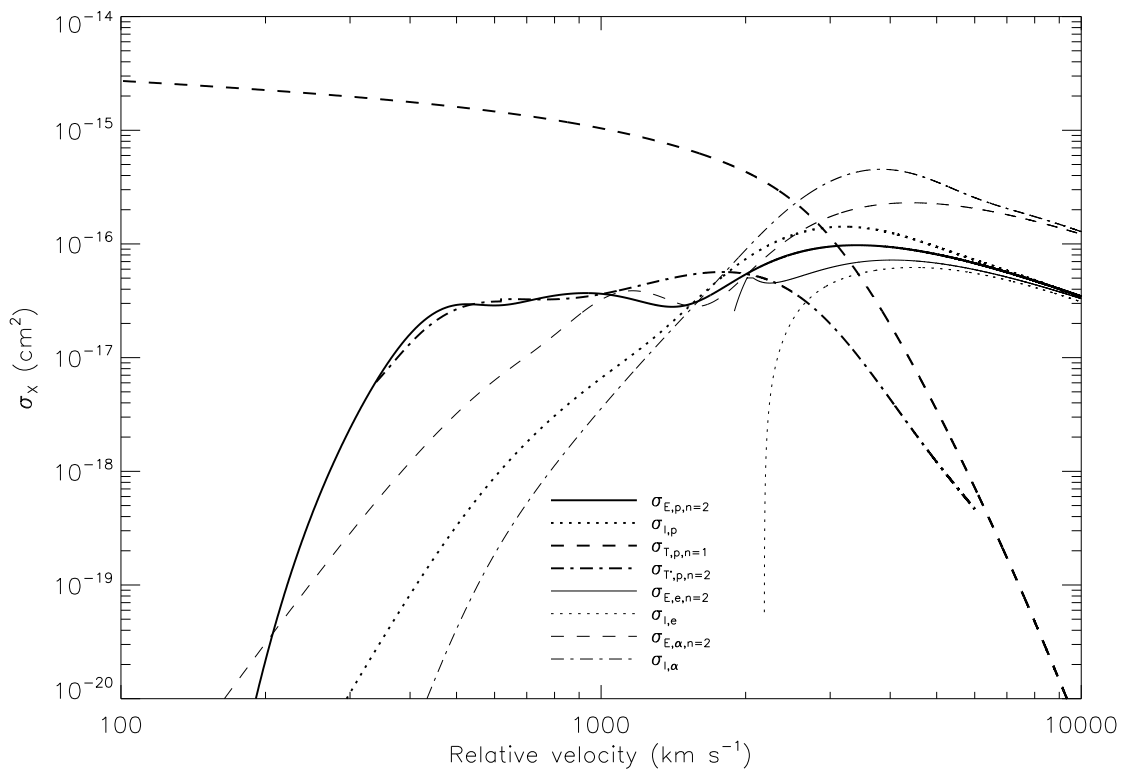}
\end{center}
\caption{Cross sections for impact excitation ($E$), ionization ($I$) and charge transfer ($T$) for interactions between neutral hydrogen atoms and various species.  The subscripts ``$p$'', ``$e$'' and ``$\alpha$'' refer to proton-atom, electron-atom and alpha particle-atom reactions, respectively.  The $E$, $I$ and $T$ ($n=1$ only) data are from Janev \& Smith (1993), while that for charge transfer to the $n=2$ state ($T^*$) are from Barnett (1990).}
\label{fig:cross}

\end{figure}

Examples of these cross sections are shown in Figure \ref{fig:cross}.  

\subsection{DISTRIBUTION FUNCTIONS AND RATE COEFFICIENTS}
\label{subsect:distrates}

\begin{figure}
\begin{center}
\includegraphics[width=6in]{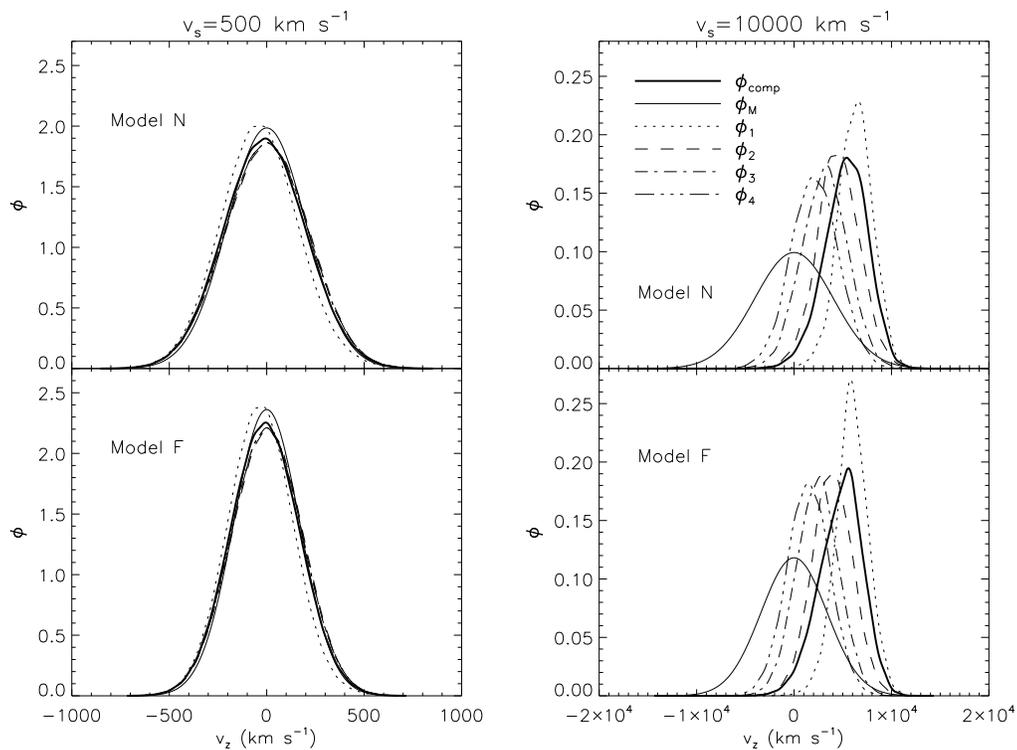}
\end{center}
\caption{Face-on projected profiles, $\phi$, for the proton and hydrogen atom velocity distributions, using arbitrary relative units.  The protons are in Maxwellian distributions ($\phi_M$), while $\phi_i$ and $\phi_{comp}$ are the atomic distributions after $i$ and one or more charge transfers (i.e., composite), respectively.  The velocity normal to the plane of the shock front is denoted by $v_z$.}
\label{fig:projdist}
\end{figure}

Figure \ref{fig:projdist} shows examples of one-dimensional velocity distribution functions of hydrogen atoms resulting from charge transfer reactions for two representative shock velocities.  The velocity axes represent the component of velocity normal to the shock surface; the distribution functions have been integrated over the transverse velocity components.  For each case, the symmetric, thin, solid curve represents the Maxwellian distribution of protons in the shocked gas, while the dotted and/or dashed curves represent the one-dimensional hydrogen atom velocity distribution functions, $\phi_i(v_z) = \int f_i(\vec{v})~dv_x dv_y$, that result from the $i$th charge transfer reactions.  The thick, solid curve represents the projected composite distribution of all hydrogen atoms in the shocked gas, $\phi_{comp}$, resulting from one or more charge transfer reactions (to be discussed in \S\ref{sect:tree}).  The two lower panels represent Model F, in which the electron and ion temperatures are fully equilibrated, while the upper panels represent Model N, in which $T_e = 0.25 T_p$ (\S\ref{sect:assumptions}).  

For shock velocities $v_s \lesssim$ 2000 km s$^{-1}$, the atomic distribution functions, $f_i$, resulting from charge transfer reactions differ only slightly from the Maxwellian distributions, $f_M$, of the protons.  We call these $f_i$ ``skewed'' Maxwellian distributions.  However, for velocities $v_s \gtrsim 2000$ km s$^{-1}$, the $f_1$ have peaks that are shifted in velocity toward the original beam, progressively more so as the shock velocity increases.  This behavior is a consequence of the fact that the charge transfer cross section decreases rapidly for relative velocities $\gtrsim$ 2000 km s$^{-1}$ (Fig. \ref{fig:cross}).  With each subsequent charge transfer, the $f_i$ shift towards $f_M$. 

\begin{figure}
\begin{center}
\includegraphics[width=6in]{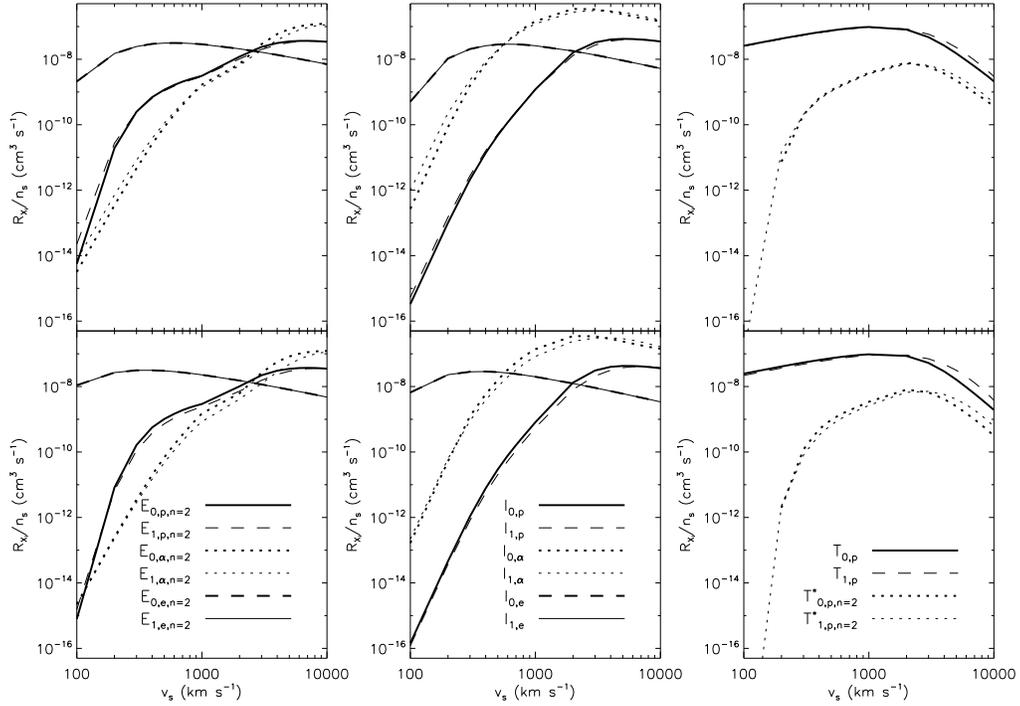}

\end{center}
\caption{Reaction rate coefficients for impact excitation ($E_i$), ionization ($I_i$) and charge transfer ($T_i$), for various species and both Models N (top row) and F (bottom row).  The subscripts ``$p$'', ``$\alpha$'' and ``$e$'' refer to proton-atom, alpha particle-atom and electron-atom reactions, respectively.  Shown are the rate coefficients for reactions involving atoms from the original beam ($i=0$) and those which are the result of one charge transfer reaction ($i=1$).}
\label{fig:rates1}
\end{figure}

Given the distribution functions and cross sections, we calculate the rate coefficients from equations (\ref{eq:rate1}) and (\ref{eq:rate2}).  These coefficients are displayed as functions of the shock velocity in Fig. \ref{fig:rates1}, for both the original beam and the atoms that result from the first charge transfer reaction.  Rate coefficients involving atoms that have undergone two or more charge transfers are nearly indistinguishable from those resulting from the first charge transfer.  Note that for shock velocities $200 \lesssim v_s \lesssim$ 2000 km s$^{-1}$, the rate coefficients for excitation, ionization and charge transfer are weakly dependent on shock velocity and are comparable in magnitude.  Note also that the excitation and ionization rate coefficients are dominated by electron collisions.  However, for shock velocities $v_s \gtrsim$ 2000 km s$^{-1}$ the net rate coefficients for excitation and ionization continue to increase, while the rates for charge transfer decrease rapidly. 

\section{THE REACTION TREE}
\label{sect:tree}

\subsection{REACTIONS PER ATOM}

\begin{figure}
\begin{center}
\includegraphics[width=6in]{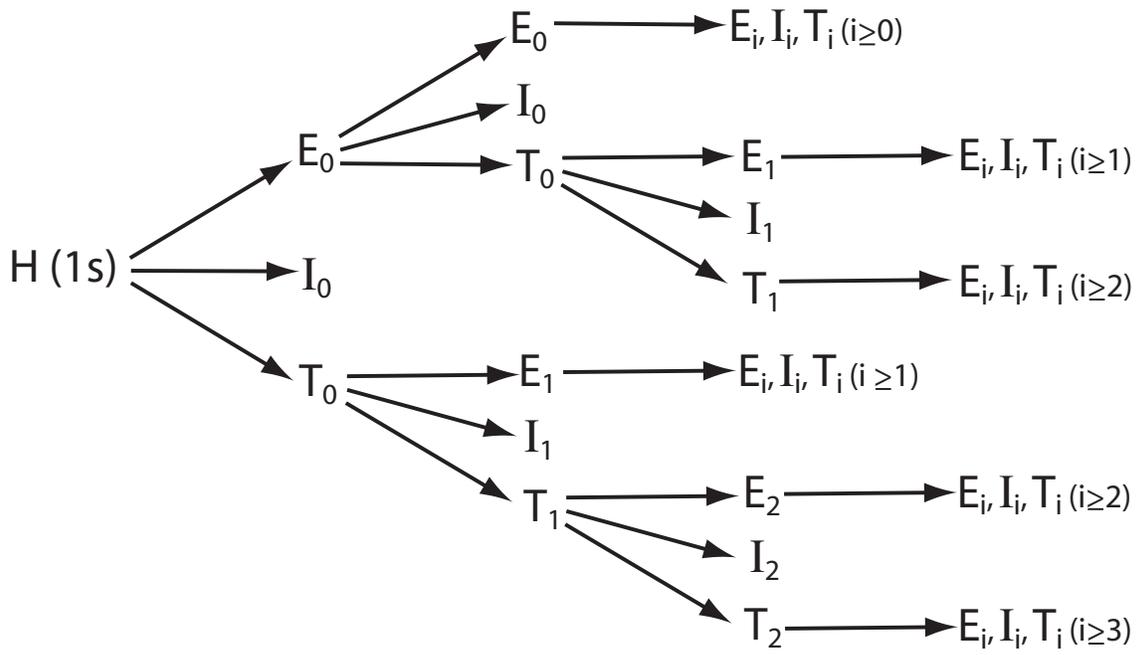}
\end{center}

\caption{Reaction tree of impact excitation ($E_i$), ionization ($I_i$) and charge transfer ($T_i$).  The index $i$ denotes the number of times a hydrogen atom has experienced charge transfer reactions.}
\label{fig:reactiontree}
\end{figure}

To calculate the ratios and profiles of hydrogen lines from a Balmer-dominated shock, we must account for all possible interactions of the hydrogen atoms.  We wish to track the fate of every hydrogen atom that crosses the shock, and to calculate the number and profile of every photon produced by the atom.  The possibilities are illustrated by the ``reaction tree'' shown in Fig. \ref{fig:reactiontree}.

Consider a hydrogen atom belonging to a distribution function, $f_i$, resulting from $i$ charge transfer reactions (or ``skewings''). The rate at which this atom will have an excitation ($E_i$) or ionization ($I_i$) is given by the sum of the rate coefficients over species (protons, electrons and alpha particles).  The rate at which the atom will have a charge transfer ($T_i$) reaction is given by $R_{T_i}= n_p \tilde{R}_{T_i}$.  Therefore, the probability that this atom will have another charge transfer reaction before it experiences an excitation or ionization is given by
\begin{equation}
P_{T_i} = \frac{R_{T_i}}{R_{E_i} + R_{I_i} + R_{T_i}},
\end{equation}
where $R_{E_i}$ is the net rate of excitations and $R_{I_i}$ is the net rate of ionizations.  

The net rate at which an atom will have an impact excitation after the $i$th charge transfer reaction is given by \begin{equation}
R_{E_i} = \sum_n ~\left[n_e\tilde{R}_{E_i,p,n} + n_p \tilde{R}_{E_i,e,n} + n_\alpha \tilde{R}_{E_i,\alpha,n}\right] \, ,
\end{equation}
where $\tilde{R}_{E_i,n}$ is the rate coefficient for excitation to the level $n$ by protons ($p$), electrons ($e$) or alpha particles ($\alpha$) (Fig. \ref{fig:rates1}).  The net probability that the atom will be excited to any level $n$ is given by\begin{equation}
P_{E_i} = \frac{R_{E_i}}{R_{E_i} + R_{I_i} + R_{T_i}}.
\end{equation}
Likewise, we define the probability, $P_{I_i}$, that the atom will have a collisional ionization after the $i$th charge transfer reaction.  

\begin{figure}
\begin{center}
\includegraphics[width=6in]{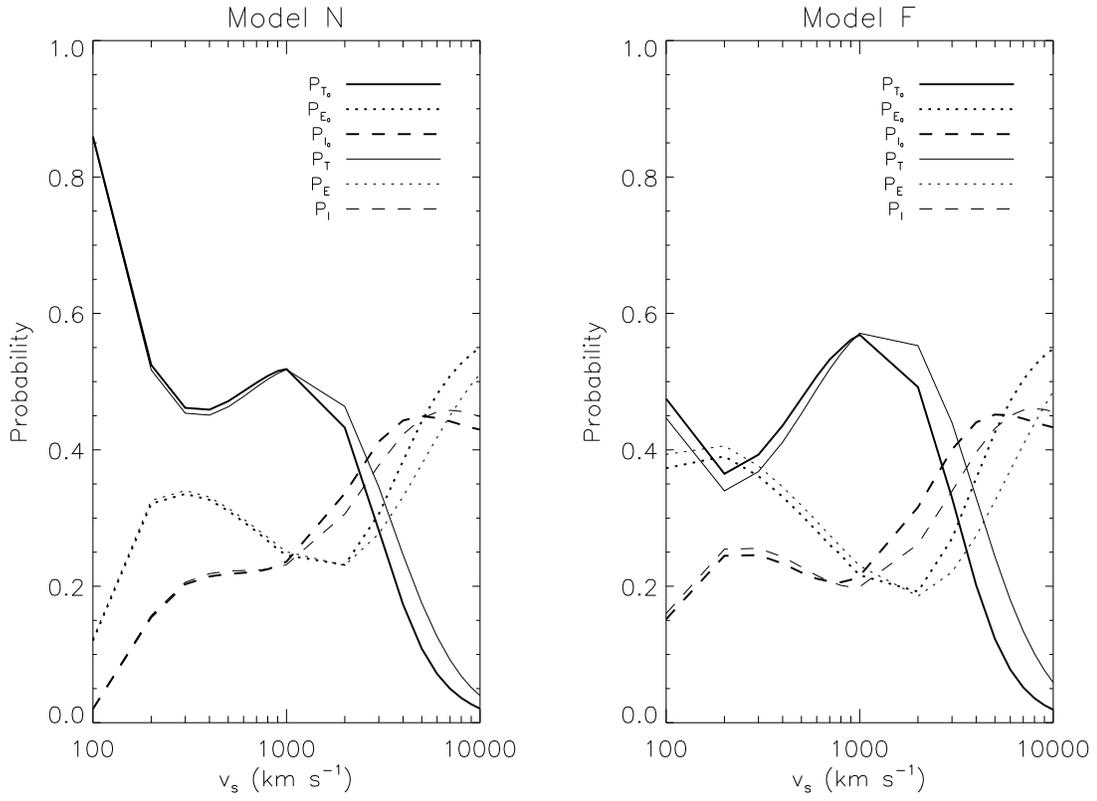}
\end{center}
\caption{Reaction probabilities, $P_{T_i}$, $P_{E_i}$ and $P_{I_i}$, for excitation, ionization, and charge transfer, respectively.  Shown are the probabilities of the reactions involving the hydrogen atom beam and the mean probabilities (see text).}
\label{fig:prob}
\end{figure}

Figure \ref{fig:prob} displays these probabilities for $i = 0$ and $i = 1$ as functions of the shock velocity.  The probabilities for $i > 1$ are almost identical to those for $i = 1$.  In practice, we average over the rate coefficients from $i = 1$ to $i = 4$ to obtain mean rate coefficients, from which we construct the probabilities.  Note that for shock velocities $v_s \lesssim 1000$ km s$^{-1}$, $P_{T_i} > P_{E_i} > P_{I_i}$.  These probabilities imply that every hydrogen atom that crosses a shock with $v_s \lesssim 1000$ km s$^{-1}$ will most likely have one or more charge transfer reactions and excitations before it is ionized.  

With these probabilities in hand, we can track a distribution of hydrogen atoms as it proceeds through the reaction tree (Fig. \ref{fig:reactiontree}).  First, consider the average number, $N_{T_0}$, of charge transfers that a hydrogen atom in the original beam will have.  Before doing so, it can have any number of excitations, so 
\begin{equation}
N_{T_0} = P_{T_0} ~\sum^\infty_{i=0} (P_{E_0})^i = \frac{P_{T_0}}{1-P_{E_0}}.
\end{equation}
The average number of ionizations by the atom in the original beam before it has one charge transfer is given by 
\begin{equation}
N_{I_0} = 1 - \frac{P_{T_0}}{1-P_{E_0}} =  \frac{P_{I_0}}{1-P_{E_0}}.
\end{equation}
Likewise, for excitations,
\begin{equation}
N_{E_0} =  \sum^\infty_{i=1} (P_{E_0})^i = \frac{P_{E_0}}{1-P_{E_0}}.
\end{equation}

\begin{figure}
\begin{center}
\includegraphics[width=6in]{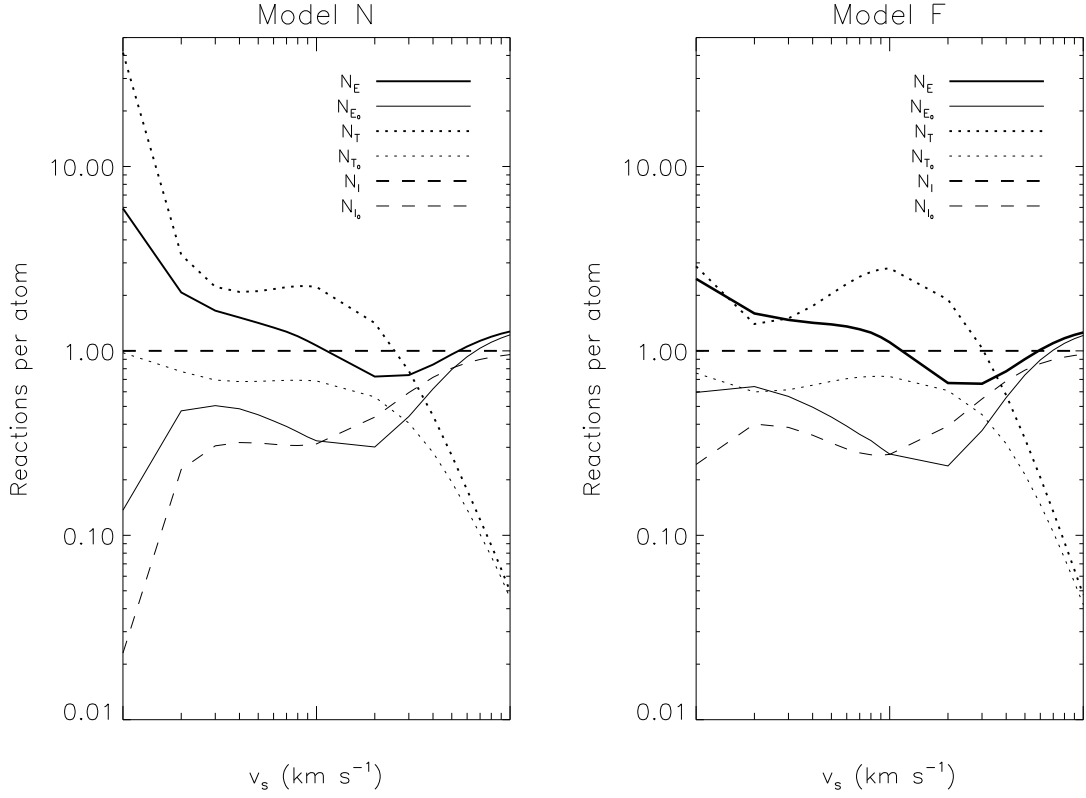}
\end{center}
\caption{Average number of excitations ($N_E$), ionizations ($N_I$) and charge transfers ($N_T$) per atom.  The zeroth order quantities are  $N_{E_0} = P_{E_0}/(1-P_{E_0})$, $N_{I_0} = P_{I_0}/(1-P_{E_0})$ and $N_{T_0} = P_{T_0}/(1-P_{E_0})$.}
\label{fig:ratios}
\end{figure}

After the first charge transfer, the probabilities, $P_{X_i}$, for excitation, ionization, and charge transfer are nearly independent of the number of charge transfers, so we may drop the subscript $i$ to describe the number of reactions per atom for the remainder of the tree.  (However, see Appendix \ref{append:tree} for the exact expressions.)  So, the average number of atoms which will survive to the $i$th charge transfer reaction is given by
\begin{equation}
N_{T_i} = \frac{P_{T_0}}{1-P_{E_0}} \left(\frac{P_T}{1-P_E}\right)^{i-1}.
\end{equation}
The average number of charge transfer reactions that will occur is given by
\begin{equation}
N_T = \frac{P_{T_0}}{1-P_{E_0}} ~\sum^{\infty}_{i=1} \left(\frac{P_T}{1-P_E}\right)^{i-1} = \frac{P_{T_0}}{P_I} \left(\frac{1-P_E}{1-P_{E_0}}\right).
\end{equation}
The average number of excitations is
\begin{equation}
N_E = \frac{1}{1-P_{E_0}} \left[ P_{E_0} + \frac{P_{T_0} P_E}{P_I} \right].
\end{equation}
Naturally, $N_I = 1$.  The number of reactions per atom, for $E$, $I$ and $T$, are plotted as a function of the shock velocity in Fig. \ref{fig:ratios}.

\subsection{LINE STRENGTHS AND PROFILES}
\label{subsect:linestrengths}

To find the intensities and profiles of emission lines, we multiply the rate at which atoms are excited by the appropriate weighting factor that a given line photon will be emitted after that reaction.  For example, for narrow line emission we find a rate per atom of 
\begin{equation}
I_{n}(\mbox{H}\alpha) = \frac{C_{32}}{1- P_{E_0}} ~\sum^m_{n=3} R_{E_0,n} ~C_{n3}
\label{eq:halpha_n}
\end{equation} 
and 
\begin{equation}
I_{n}(\mbox{Ly}\alpha) = \frac{1}{1- P_{E_0}} ~\sum^m_{n=2} ~ R_{E_0,n} ~C_{n2},
\end{equation} 
where the $C_{nn^\prime}$ are the cascade matrices (Seaton 1959) representing the probability that a hydrogen atom excited to state $n$ will make a transition to the state $n^\prime<n$ via all cascade routes. In Appendix \ref{append:cascade}, we present analytical expressions for  $C_{nn^\prime}$ and note that excitations are not necessarily distributed statistically among the angular momentum states.  The quantity $R_{E_0,n}$ is the rate coefficient for excitations to the state $n$ before any charge transfer reaction occurs.

The broad line emission has contributions from all excitations after the first charge transfer and also from charge transfer reactions to excited states:
\begin{equation}
I_b(\mbox{H}\alpha) = \frac{C_{32}}{1-P_{E_0}} ~\sum^m_{n=3} ~\left[\frac{P_{T_0}}{P_I} ~\left(R_{E,n} + R_{T^*,n} \right) + R_{T^*_0,n} \right]~C_{n3}
\label{eq:halpha_b}
\end{equation}
and
\begin{equation}
I_b(\mbox{Ly}\alpha) = \frac{1}{1-P_{E_0}} ~\sum^m_{n=2} ~\left[\frac{P_{T_0}}{P_I} ~\left(R_{E,n} +  R_{T^*,n} \right) + R_{T^*_0,n} \right]~C_{n2},
\end{equation}
where $R_{E,n}$ is the rate of excitations to state $n$ after any number $i \ge 1$ of charge transfer reactions and $R_{T^*,n}$ is the corresponding rate of charge transfer reactions to excited states.  

We construct a composite distribution function, $f_{comp}$, for the hydrogen atom velocity distribution by summing the $f_i$, weighted by the probabilities that the atoms undergo $i$ charge transfers:
\begin{equation}
f_{comp}(\vec{v}) \propto \sum^{\infty}_{i=1} ~\left[\frac{P_T}{1-P_E}\right]^{i-1} ~f_i(\vec{v}).
\label{eq:fcomp}
\end{equation}
Figure \ref{fig:projdist} shows the function $\phi_{comp}$, which is $f_{comp}$ integrated over velocities transverse to the shock, for two representative shock velocities.  Note that for $v_s = 500$ km s$^{-1}$, $\phi_{comp}$ differs only slightly from the projected proton distribution, $\phi_M$, hence validating the assumption made in earlier models.  In contrast, for $v_s = 10,000$ km s$^{-1}$, $\phi_{comp}$ differs greatly from $\phi_M$.  Its peak is shifted substantially toward the velocity of the original beam.  In fact, it is dominated by $\phi_1$.  It contains a small contribution from $\phi_2$ and negligible contributions from $\phi_{i>2}$, owing to the small probabilities that the atoms will undergo multiple charge transfer reactions.

For a face-on shock, the velocity profile of the broad line is given by $\phi_{comp}$.  Since $f_{comp}$ is not isotropic, the line profile will depend on the aspect (or viewing) angle, $\theta_v$.  To investigate this angle dependence, we perform a coordinate transformation from cylindrical to Cartesian coordinates and then rotate the distribution about the $x$-axis by $\theta_v$.  Figure \ref{fig:fwhm1} shows the full width at half-maximum (FWHM) of the broad emission line as a function of shock velocity for shocks viewed face-on ($\theta_v=0$) and edge-wise ($\theta_v=90^\circ$).  Because the ions are hotter in the non-equilibrated Model N than in the fully-equilibrated Model F, we see that the FWHM is greater (by a factor of between 1.1 and 1.2) in model N than in Model F.  We also see that the FWHM increases almost linearly with shock velocity for $v_s \lesssim 2000$ km s$^{-1}$ but that it begins to level off for $v_s \gtrsim 2000$ km s$^{-1}$, consistent with earlier studies.  In general, the FWHM is a weak function of $\theta_v$; it is slightly greater (by a factor $\approx 1.1$) for edge-wise than for face-on shocks.

\section{RESULTS}

\label{sect:results}

The observable quantities for Balmer-dominated shocks are the profiles (or widths) of the broad lines, the velocity shifts between the peaks of the broad line and narrow lines, and the intensities of the broad and narrow lines.  Below, in \S\ref{subsect:snr}, we discuss the interpretation of these quantities for several well-known SNRs in the light of the models described above and we compare our results to previous work.  Then, in \S\ref{subsect:snr1987a}, we discuss how our results may be used to interpret the emission from the reverse shock in SN 1987A.

\subsection{BALMER-DOMINATED REMNANTS}
\label{subsect:snr}

\begin{figure}
\begin{center}
\includegraphics[width=6in]{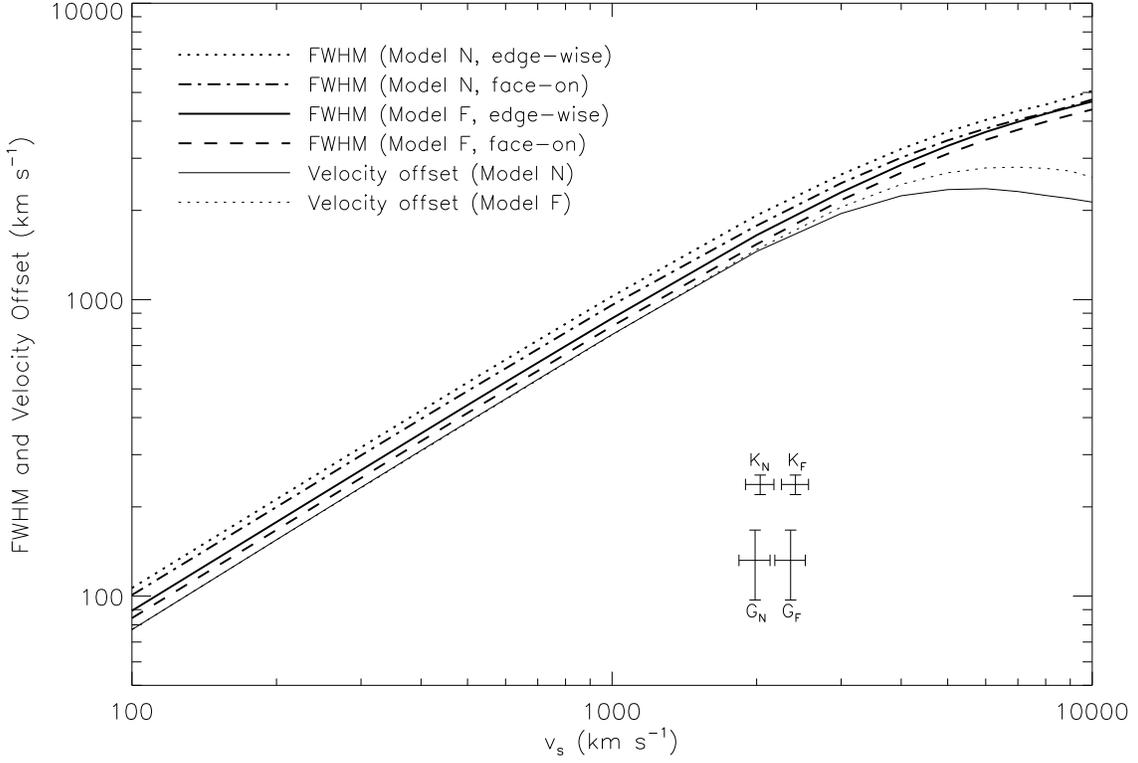}
\end{center}
\caption{Full width at half-maximum (FWHM) of the projected profiles of $f_{comp}$.  These are shown for both face-on and edge-wise shock fronts.  Also shown are the velocity offsets between the peaks of the narrow and the face-on broad components ($\phi_{comp}$).  The data points represent velocity offsets measured for Tycho by Kirshner, Winkler \& Chevalier (1987; $238 \pm 18$ km s$^{-1}$) and Ghavamian et al. (2001; $132 \pm 35$ km s$^{-1}$), which are denoted by ($K_N$, $K_F$) and ($G_N$, $G_F$), respectively; the subscripts stand for Models N and F.  We assume a face-on shock front to derive the shock velocity range for each data point.}
\label{fig:fwhm1}
\end{figure}

As CKR80 and others have emphasized, we can infer the velocity, $v_s$, of a Balmer-dominated shock from the observed FWHM.  Our results are fairly close to previous ones.  For example, for our fully equilibrated Model F (viewed edge-wise), and $v_s = 2000$ km s$^{-1}$, we find FWHM $= 1600$ km s$^{-1}$ (Fig. \ref{fig:fwhm1}), which may be compared to corresponding values in the literature: FWHM $\approx 1800$ km s$^{-1}$ estimated from Fig. 3 of CKR80, FWHM $\approx 1600$ km s$^{-1}$ estimated from Fig. 7 of S91, and FWHM $\approx 1400$ km s$^{-1}$ estimated from Fig. 9 of G01. Unfortunately, we have not been able identify the sources for the differences among these values.  However, we note that there is a typographical error in equation (5) of CKR80, which propagated into papers like KWC87; this was later corrected in S91.

In Table 1 (column 3), we list values of the FWHM of the broad H$\alpha$ line measured in several SNRs.  Note that values measured by different authors for a given SNR do not necessarily apply to the same part of the shock front, so the differences among such measurements may be real.  We also list in column 4 the range of shock velocities inferred by those authors by fitting their models to the observations.  This range takes into account uncertainties both in the measured FWHM and in the model itself, with the lower values of $v_s$ coming from non-equilibrated models and the higher values from fully equilibrated ones.  In columns 6 and 7, we list the values of $v_s$ that we infer by fitting our models to the same data.  The values in column 6 are inferred from the non-equilibrated Model N, while those in column 7 are from the fully equilibrated Model F, and the errors in each case derive only from the errors in the measured values of the FWHM.  (In making such fits, we used the model curve for edge-wise aspect, which we believe to be close to the actual situation in most cases.) In general, we see that the values of $v_s$ inferred from our models are in fairly good agreement with those inferred by other authors, but that in some cases we infer values of $v_s$ that are lower than the original values by as much as 28\%.

The velocity offset, $\Delta v(v_s,\theta_v)$, between the centroids of the narrow and broad components is sensitive both to the shock velocity, $v_s$, and the aspect angle, $\theta_v$, of the shock.  We can write 
\begin{equation}
\Delta v(v_s,\theta_v) = \Delta v(v_s,0) ~\cos(\theta_v),
\end{equation}
where $\Delta v(v_s, 0)$, the offset for a face-on shock, is shown in Fig. \ref{fig:fwhm1}.  At low shock velocities, the composite profile is virtually coincident with the Maxwellian of the protons and so, to a very good approximation, $\Delta v(v_s,0) = 3v_s/4$.  However, for $v_s \gtrsim 2000$ km s$^{-1}$, the composite profiles (Fig. \ref{fig:projdist}) begin to shift noticeably toward the narrow line, so that $\Delta v(v_s,0) < 3v_s/4$. 

It should be possible to infer the aspect angle of the shock by comparing the measured FWHM with $\Delta v$.  For example, KWC87 measured FWHM $= 1800 \pm 100$ km s$^{-1}$ and $\Delta v(v_s,\theta_v) = 238 \pm 18$ km s$^{-1}$ in a filament in Tycho's remnant.  From this result and Fig. \ref{fig:fwhm1}, we can infer $v_s = 2409 \pm 156$ km s$^{-1}$ and $\theta_v = 84^{+6.0^\circ}_{-8.4^\circ}$ for Model F, consistent with KWC87's estimation of $90^\circ - \theta_v \approx 6^\circ$.  Of course, the observations will tend to select bright filaments that are viewed nearly edge-on.

The measured ratio of broad to narrow line H$\alpha$ intensities, $\Re_{b,n}(\mbox{H}\alpha) = I_b(\mbox{H}\alpha)/ I_b(\mbox{H}\alpha)$, is listed in column 5 of Table 1.  As first noted by CKR80, this ratio is sensitive to the optical depth in the in the Lyman series.  In choosing the cascade matrices in equations (\ref{eq:halpha_n}) and (\ref{eq:halpha_b}), we consider two limits.  In Case A, all Lyman transitions are optically thin.  In Case B, they are optically thick; therefore, any Ly$\beta$ photon that is emitted will be trapped until it splits into Ly$\alpha +$H$\alpha$.  Likewise, Ly$\gamma$ must be converted to Ly$\alpha +$H$\beta$ or Ly$\alpha +$H$\alpha + $P$\alpha$, and so on.  In Balmer-dominated shocks, the narrow line emissivity of H$\alpha$ for Case B is increased by a factor $\approx 2.5$ compared to Case A, and that of H$\beta$ by a factor $\approx 1.9$.  G01 modeled the ionization structure of the shock fronts in three SNRs and performed Monte Carlo simulations of Ly$\beta$ and Ly$\gamma$ trapping.  They find that most of the broad Lyman photons escape from the shock, but that narrow Lyman photons are likely to be absorbed and converted into Balmer lines both behind and ahead of the shock.  Therefore, for most SNRs (except SNR 1987A), we are making a good approximation to adopt Case A for the broad lines and Case B for the narrow lines.  For a theoretical upper limit on $I_b/I_n$, we assume Case A for both components.  

\begin{figure}
\begin{center}
\includegraphics[width=6in]{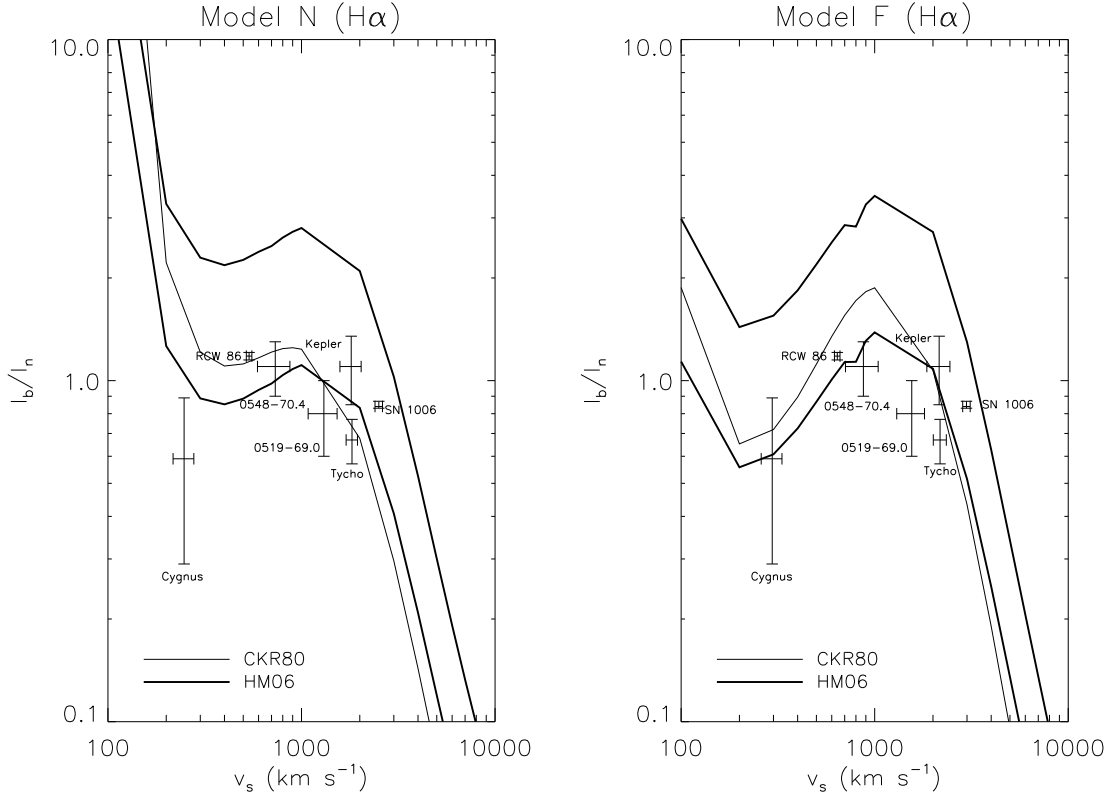}
\end{center}
\caption{Ratio of the broad to narrow H$\alpha$ emission, $\Re_{b,n}(\mbox{H}\alpha)$, {\it vs.} shock velocity, $v_s$.  Each pair of thick curves represents our lower and upper bounds for $\Re_{b,n}$ (denoted by the label ``HM06'').  The ``CKR80'' curve is obtained using the expression of Chevalier, Kirshner \& Raymond (1980) and Smith et al. (1991).  The data represent SN 1006 (Ghavamian et al. 2002), Kepler (Fesen et al. 1989), 0519-69.0, 0548-70.4 (Smith et al. 1991), Tycho, RCW 86 and Cygnus (Ghavamian et al. 2001), assuming edge-wise shocks.}
\label{fig:halpha}
\end{figure}

In Figure \ref{fig:halpha}, we plot the measurements of $\Re_{b,n}(\mbox{H}\alpha)$ and $v_s$ inferred from the measured FWHM for several SNRs.  We do this for both the non-equilibrated Model N and the fully equilibrated Model F. As in columns 6 and 7 of Table 1, the uncertainties in the inferred $v_s$ are due entirely to the measurement errors.  We also plot, for both models, the relationship $\Re_{b,n}(\mbox{H}\alpha)$ {\it vs.} $v_s$ predicted by our theory, as well as the same relationship from equation (4) of S91 (using our rate coefficients), 
\begin{equation}
\Re_{b,n}(\mbox{H}\alpha) = \frac{R_{T_0}}{\epsilon_B R_I} \left[ \epsilon_A + g_\alpha \left( 1 + \frac{R_T}{R_I} \right) \right],
\end{equation}
which follows from the model originally presented by CKR80. The Case A and B efficiencies ($\epsilon_A$ and $\epsilon_B$) are given in CR78, while we adopt the mean value of the fraction of charge transfers to excited states yielding H$\alpha$ photons to be $g_\alpha \approx 0.03$.  In both cases, we assume Case A for $I_b$ and Case B for $I_n$.

One can see from Figure \ref{fig:halpha} that the data fit the models fairly well, and that our derived relationship of $\Re_{b,n}(\mbox{H}\alpha)$ {\it vs.} $v_s$ does not differ very much from that derived by CKR.  In the cases of SN 1006, Kepler and RCW 86 and 0548---70.4, the fits do not discriminate very well between Models N and F, though Model F provides a better fit to the Cygnus Loop than Model N.  The remnants of Tycho and 0519-69.0 are marginally fitted by Models F and N, respectively.  With the exception of SN 1006, we also note that the data cannot be fit with a model in which Case A applies to $I_n$ as well as $I_b$.  In that case, the theoretical curves would be elevated by a factor $\approx 2.5$ (Fig. \ref{fig:halpha}), and would not fit the data at all.

\begin{figure}
\begin{center}
\includegraphics[width=6in]{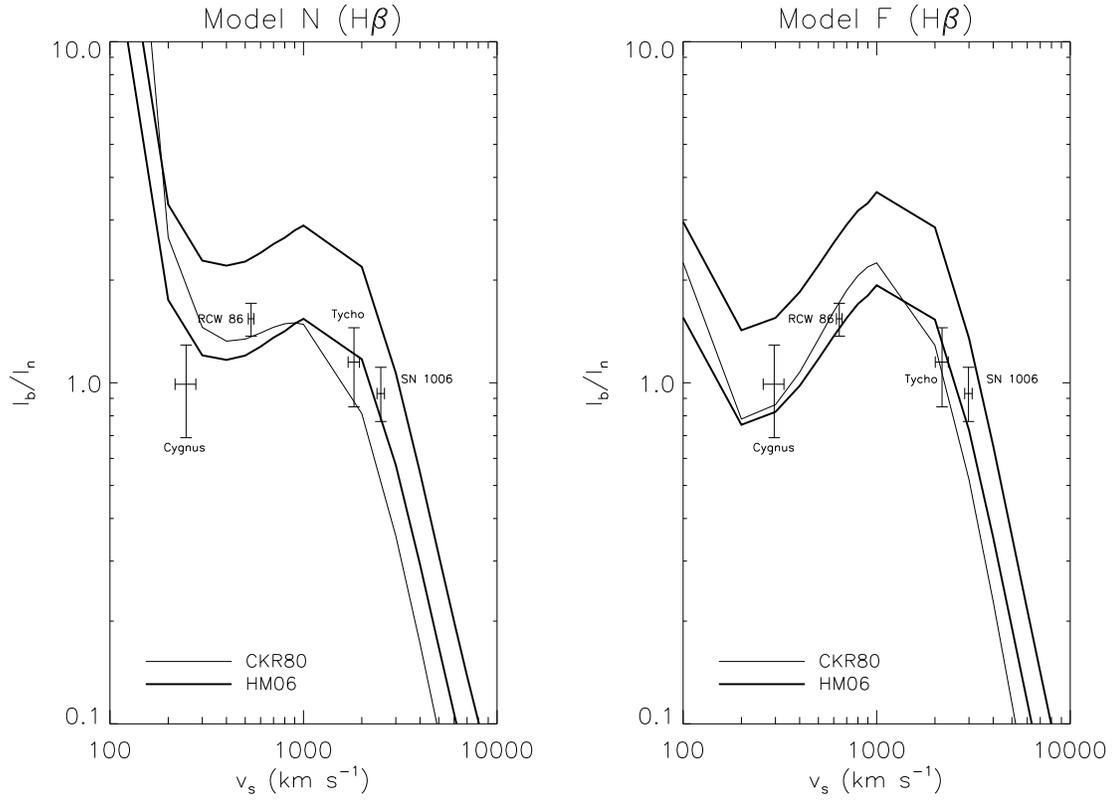}
\end{center}

\caption{Same as Fig. \ref{fig:halpha}, but for H$\beta$ emission.  The data represent SN 1006 (Ghavamian et al. 2002), Tycho, RCW 86 and Cygnus (Ghavamian et al. 2001), again assuming edge-wise shocks.}
\label{fig:hbeta}
\end{figure}

We can make a similar comparison of theory to data with the H$\beta$ line.  The corresponding fraction of charge transfer to excited states yielding H$\beta$ is estimated to be $g_\beta \approx 0.01$ (Bates \& Dalgarno 1953).  Figure \ref{fig:hbeta} shows both the measured and theoretical values of $\Re_{b,n}(\mbox{H}\beta)$ {\it vs.} $v_s$.  Again, the difference between our model and that of CKR80 is not great.  The fit of the theory to the data does not discriminate between Models N and F, except for the Cygnus Loop where the latter is favored.

Note that the ratios $\Re_{b,n}(\mbox{H}\alpha)$ and $\Re_{b,n}(\mbox{H}\beta)$ are fairly insensitive to shock velocity in the range $200 \lesssim v_s \lesssim 2000$ km s$^{-1}$.  However, for $v_s \gtrsim 2000$ km s$^{-1}$, these ratios do become quite sensitive to $v_s$.  It is worthwhile to note that the Monte Carlo simulations of G01 and G02, which model Lyman line trapping, result in tighter constraints on $\Re_{b,n}$.

\subsection{SNR 1987A}
\label{subsect:snr1987a}

SNR 1987A has a double shock structure, consisting of a forward and a reverse shock.  In contrast to other SNRs, the Balmer-dominated shock in SNR 1987A comes, not from nearly stationary neutral hydrogen overtaken by the supernova blast wave, but from freely-expanding neutral hydrogen atoms in the supernova debris crossing the reverse shock.   The debris crosses the reverse shock at velocities $\sim$ 12,000 km s$^{-1}$ (M03; H06).  In the frame of the observer, the forward shock or blast wave is moving at between 3500 and 5200 km s$^{-1}$ (Manchester et al. 2002; Park et al. 2002; Michael et al. 2002).   The reverse shock is moving at about 80\% of the velocity of the forward shock (M03).  Hence, in the frame of the observer, the post-shock ions are relatively slow particles moving at velocities of between 5100 and 6120 km s$^{-1}$.  The ``shock velocity'' in this case is interpreted as the velocity of the freely-streaming hydrogen atoms in the rest frame of the reverse shock, i.e., $7840 \leq v_s \leq 9200$ km s$^{-1}$.

In the terminology of this paper, the ``narrow'' line emission resulting from excitation of hydrogen atoms initially crossing the reverse shock will have the greatest Doppler shift, corresponding to the free-streaming debris.  There is a unique mapping of the Doppler shift of this narrow emission to the depth of the reverse shock measured from the midplane of the supernova debris along the line of sight.  Using the Space Telescope Imaging Spectrograph (STIS), M03 and H06 (and references therein) mapped streaks of high velocity emission in both H$\alpha$ and Ly$\alpha$ in SNR 1987A, which H06 called ``surface emission''.  This surface emission is evidently equivalent to narrow line emission from atoms crossing the reverse shock.  H06 also detected H$\alpha$ and Ly$\alpha$ emission from the same location having substantially lower Doppler shifts than the surface emission.  If that emission came from freely-expanding atoms in the supernova debris, those atoms would have to reside beneath the reverse shock surface.  Accordingly, H06 called this emission ``interior emission''.  That was an unfortunate choice of terminology.  At least part if not all of this emission must come from more slowly moving hydrogen atoms immediately {\it outside} the reverse shock surface, which result from charge exchange reactions of hydrogen atoms in the supernova debris with protons in the shocked gas.  This ``interior emission'' is analogous to the broad line emission in the terminology of this paper.  

Following M03, we have adopted a model for SNR 1987A having $\chi_{He} = n_{He}/n_H = 0.2$.  As before, we consider two variants, Model N ($T_e/T_p = 0.25$, $f_{eq} \approx 0.33$) and the fully equilibrated Model F ($T_e/T_p = 1$).  In SNR 1987A, the debris crossing the reverse shock has a Sobolev optical depth $\tau_{Ly\alpha} \sim 1000$ (M03).  Therefore, we assume that Case B is a good approximation, not only for the narrow line emission, but also for the broad line emission. 

Since the broad and the narrow line emission from SNR1987A comes from the expanding debris as it crosses the reverse shock, the line profiles will depend on the hydrodynamics of the reverse shock surface, which we do not consider here.  However, one can measure the ratios of broad (``interior'') to narrow (``surface'') emission and of H$\alpha$ to Ly$\alpha$ emission (H06).  By comparing those ratios to those predicted by our theory, we can test the hypothesis that the interior emission is indeed the result of charge transfer reactions between hydrogen atoms and protons at the reverse shock.  

\begin{figure}
\begin{center}
\includegraphics[width=6in]{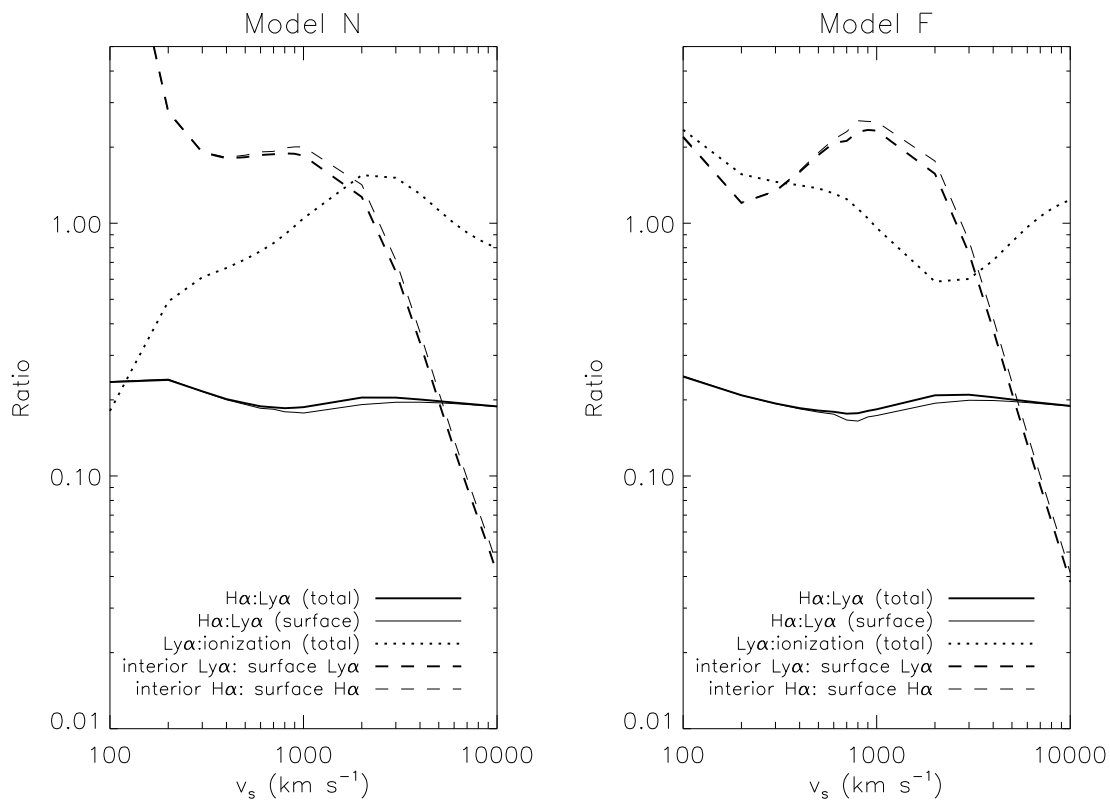}
\end{center}

\caption{Various ratios of the Ly$\alpha$ to H$\alpha$ and Ly$\alpha$ to ionization rate coefficients for conditions relevant to SNR 1987A (Case B and 20\% alpha particles by number compared to hydrogen).  ``Total'' refers to the sum of surface and interior emission, as defined by Heng at al. (2006; see text).}
\label{fig:87a}
\end{figure}

Figure \ref{fig:87a} shows these ratios as functions of shock velocity for our model of SNR 1987A.  The solid curves show the photon emission ratio $\Re_{HL} = I(\mbox{H}\alpha)/I(\mbox{Ly}\alpha)$.  We see that $\Re_{HL} \approx 0.19$, a result that is almost independent of $v_s$ and almost the same for interior and surface emission, both for Model N and Model F.  This value is consistent with the value $\Re_{HL} \approx 0.2$ estimated by CKR80 and $\Re_{HL} \approx 0.21$ estimated by M03.

The dotted curves in Fig. \ref{fig:87a} show the number, $N_{Ly\alpha}$, of Ly$\alpha$ photons (both surface and interior) emitted per hydrogen atom crossing the reverse shock.  In the rest frame of the reverse shock, these atoms cross the shock with a velocity $7840 \leq v_s \leq 9200$ km s$^{-1}$.  For $v_s$ in this range, $N_{Ly\alpha} \approx 1.1$ for both Models N and F. These values may be compared to the value $N_{Ly\alpha} \approx 1$ estimated by M03.  

The dashed curves in Fig. \ref{fig:87a} show the ratios $\Re_{b,n}(\mbox{H}\alpha)$ of interior to surface emission.  For the range of shock velocity expected in SNR 1987A, the curves predict $\Re_{b,n}(\mbox{H}\alpha) \approx 0.08$ for both Models N and F.  These values are more than an order of magnitude smaller than the ratio $\Re_{b,n}(\mbox{H}\alpha) \sim 1$ observed by H06.

\section{DISCUSSION}
\label{sect:discussion}

For most SNRs, the models presented here agree fairly well with previous models and seem to account for the observed data of both broad line FWHM and ratio $\Re_{b,n}(\mbox{H}\alpha) \approx 1$ of broad-to-narrow line intensities.  However, these same models fail to account for the observed ratio $\Re_{b,n}(\mbox{H}\alpha)$ of interior-to-surface emission for the conditions of the reverse shock in SNR 1987A.  Either the model is wrong, or the interior H$\alpha$ emission from SNR~1987A is dominated by some excitation mechanism other than charge transfer at the reverse shock.  

As we mentioned in \S\ref{sect:assumptions}, the most questionable assumption of the model is that the hydrogen atoms enter a proton plasma which has been suddenly decelerated according to shock jump conditions.  This assumption is flawed --- the approximation is good only when the neutral fraction is small (see Lim \& Raga 1996 for a discussion on the effects of larger neutral fractions).  Indeed, the reverse shock has a transition zone of finite thickness $\Delta z \sim n_p \sigma_{I,p}$, within which the ionization, excitation and charge transfer reactions occur.  Consider the first hydrogen atoms to enter this zone from upstream.  When they become ionized, the resulting protons must have streaming velocities comparable to that of the hydrogen atoms.  Within the transition zone, the newly created protons must be decelerated and compressed so that they emerge on the downstream side of the zone with velocity $v_p = v_s/4$.  But most of the ionizations must occur when the relative streaming velocity of the hydrogen atoms and the protons is substantially less than the relative velocity of $3v_s/4$ that is assumed in the present model.  Given the fact that $\Re_{b,n}(\mbox{H}\alpha)$ decreases precipitously for $v_s \gtrsim 2000$ km s$^{-1}$ (Fig. \ref{fig:87a}), we see that our model might grossly underestimate the actual value of this ratio in the case of SNR 1987A.  

To properly interpret the H$\alpha$ and Ly$\alpha$ emission from SNR 1987A, we must construct a model for this transition zone similar to that of Whitney \& Skalafuris (1963) and Skalafuris (1965, 1968 and 1969).  In a sense, the analysis presented here is a warm-up to that task.  Once we have such a model, we can use the results of the present analysis to calculate the emissivity and profiles of broad and narrow lines throughout the transition zone. 

The assumption that a Balmer-dominated shock can be treated as a discontinuous jump is faulty for all SNRs, not just SNR~1987A.  Accordingly, we may ask why the model seems to work for most SNRs, for which $v_s \lesssim 2000$ km~s$^{-1}$.  If one allowed for the transition zone in constructing a model for actual SNRs, one would construct a weighted average of $\Re_{b,n}(\mbox{H}\alpha)$ over relative velocities ranging up to $3v_s/4$.  But, as Fig. \ref{fig:halpha} shows, the ratio $\Re_{b,n}(\mbox{H}\alpha)$ does not vary greatly for $200 \lesssim v_s \lesssim 2000$ km~s$^{-1}$.  Therefore, the value of $\Re_{b,n}(\mbox{H}\alpha)$ appropriately averaged over a range of relative velocities would still agree fairly well with the observed values.  

On the other hand, these considerations bring into question the validity of the relationship between the broad line FWHM and the shock velocity predicted by the current models.  In a more realistic model that takes into account the finite thickness of the transition zone, a greater fraction of the H$\alpha$ emission would take place where the relative velocity is less than $3v_s/4$.  The result would be a reduced value of the FWHM of the broad emission for a given value of $v_s$ --- or, equivalently, a greater value of $v_s$ required to account for a given FWHM.  This consideration suggests that the current models may be under-estimating significantly the shock velocities of SNRs.

The present model is also inadequate to interpret current observations of the ratio $\Re_{b,n}(Ly\alpha)$ in SNR 1987A.  The observed brightness of Ly$\alpha$ is modified by resonant scattering in the supernova debris.  The resonance scattering can double the brightness of Ly$\alpha$ on the near side of the supernova debris and greatly suppress it on the far side, as pointed out by H06. However, since the Ly$\alpha$ photons are produced in the transition zone, and their mean free paths are less than $\Delta z$, we need a model of the transition zone to account properly for this resonant scattering.

\acknowledgments \scriptsize

K.H. wishes to thank following people: Lin Chii-Dong and David Schultz for invaluable advice regarding the reaction cross sections; Roger Chevalier, Svetozar Zhekov, J. Michael Shull and Andrew Hamilton for illuminating discussions;  Davide Lazzati and Dmitri Veras for help on technical issues; Brian Morsony for his careful reading of the manuscript.  He is deeply grateful to James Green (through NASA Grant NAG5-12279) and Ropert Kirshner for financial support.  He also acknowledges the expertise of James McKown and Peter Ruprecht, who provided impeccable technical support at JILA, without which this piece of work would not have been possible.  We thank the anonymous referee for helpful comments which improved the manuscript.

\normalsize


\appendix \normalsize
\section{APPENDIX}

\subsection{ADDITIONAL EQUATIONS FOR THE REACTION TREE}
\label{append:tree}

Consider a hydrogen atom in the distribution function $f_0$.  If it undergoes an infinite number of charge transfers, the average number of excitations it will have is
\begin{equation}
N_{E} = \frac{P_{E_0}}{1-P_{E_0}} + \sum_{j=0}^\infty \prod_{k=0}^j ~\frac{P_{T_k}}{1-P_{E_k}} ~\frac{P_{E_{j+1}}}{1-P_{E_{j+1}}},
\label{eq:ne}
\end{equation}
where the second term accounts for excitations after one or more charge transfers.  In the case of a reaction tree with no $T_i$ reactions, equation (\ref{eq:ne}) simply reduces to the familiar $P_{E_0}/(1-P_{E_0})$.  Similarly, the average number of ionizations and charge transfers per atom are
\begin{equation}
N_{I} = \frac{P_{I_0}}{1-P_{E_0}} + \sum_{j=0}^\infty \prod_{k=0}^j ~\frac{P_{T_k}}{1-P_{E_k}} ~\frac{P_{I_{j+1}}}{1-P_{E_{j+1}}}
\end{equation}
and
\begin{equation}
N_{T} = \sum_{j=0}^\infty \prod_{k=0}^j \frac{P_{T_k}}{1-P_{E_k}},
\end{equation}
respectively.  

\begin{figure}
\begin{center}
\includegraphics[width=6in]{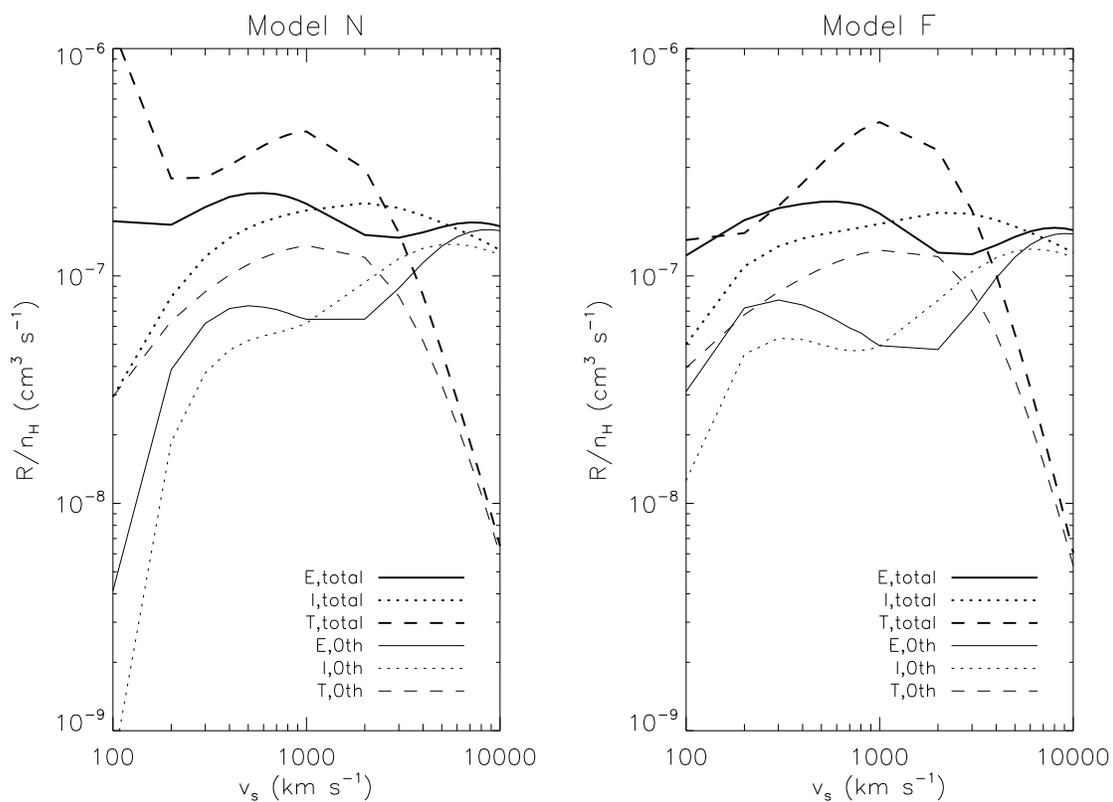}
\end{center}
\caption{The total rate coefficient, $R_{X,total}/n_H$, weighted by the fractional abundance of each species (protons, electrons and alpha particles), for excitation, ionization and charge transfer.  The zeroth order counterparts (see \S\ref{append:tree}) are denoted by ``0th''.}
\label{fig:totalrates}
\end{figure}

The total rate is obtained in an analogous way:
\begin{equation}
R_{X,total}= \frac{R_{X_0}}{1-P_{E_0}} + \sum_{j=0}^\infty \prod_{k=0}^j ~\frac{P_{T_k}}{1-P_{E_k}} ~\frac{R_{X_{j+1}}}{1-P_{E_{j+1}}}.
\end{equation}
If we make the approximation that $R_{X_i} = R_X$ for $i \geq 1$, then the preceding expression reduces to:
\begin{equation}
R_{X,total} = \frac{1}{1-P_{E_0}} \left( R_{X_0} + \frac{P_{T_0}}{P_I} ~R_X\right).
\end{equation}
The zeroth order rate is then $R_{X_0}/(1-P_{E_0})$.  Examples of $R_{X_0}/n_H(1-P_{E_0})$ and $R_{X,total}/n_H$ are plotted in Fig. \ref{fig:totalrates}.

\subsection{CASCADE MATRIX}
\label{append:cascade}

Consider the collisional excitation of a hydrogen atom from the ground state to the level $n$.  Let $P_{nn'}$  be the probability that the population of $n$ is followed by a direct radiative transition to $n'$ (Seaton 1959), given by
\begin{equation}
P_{nn'} = \frac{A_{nn'}}{A_n},
\end{equation}
where
\begin{equation}
A_n = \sum_{i=n_0}^{n-1} A_{ni},
\end{equation}
and we have $n_0$=1 for Case A and $n_0$=2 for Case B.

Let $C_{nn'}$ be the probability that the population of $n$ is followed by a transition to $n'$ via {\it all} possible cascade routes.  Following Seaton (1959), the cascade matrix elements are
\begin{equation}
C_{nn'} = \sum_{i=n'}^{n-1} P_{ni} C_{in'},
\label{eq:seaton}
\end{equation}
where, for any $a \geq b$, $C_{ab}$=1 if $a=b$.  An analytically more illuminating form is
\begin{eqnarray}
C_{nn'} = P_{nn'} + \sum_{i_1=n'+1}^{n-1} P_{ni_1} P_{i_1n'} + \sum_{i_1=n'+2}^{n-1} \sum_{i_2=n'+1}^{n-2} P_{ni_1} P_{i_1i_2} P_{i_2n} \nonumber\\
+ \sum_{i_1=n-2}^{n-1} \sum_{i_2=n-2}^{n-3} ... \sum_{i_{N-1}=n+2}^{n+3} \sum_{i_N=n+1}^{n+2} P_{ni_1} P_{i_1i_2} ... P_{i_{N-1}i_N} P_{i_Nn'} \nonumber\\
+ P_{n,n-1} P_{n-1,n-2} ... P_{n+2,n+1} P_{n+1,n},
\end{eqnarray}
where for any $a \geq b$, $P_{a,b} \equiv P_{ab}$ and $P_{ab}$=1 for $a=b$.

Determining the cascade matrix elements requires knowing the Einstein A-coefficients.  In the electric dipole approximation, they are (Pengelly 1964)
\begin{equation}
A_{nn'} = \frac{1}{n^2} \sum_{l'=0}^{n-1} \sum_{l=l'\pm1} (2l+1) A_{nl,n'l'},
\end{equation}
where the quantity $A_{nl,n'l'}$ is given by (e.g., Pengelly 1964; Brocklehurst 1971)
\begin{equation}
A_{nl,n'l'} = 2.6774 \times 10^9 \left( \frac{1}{n^{'2}} - \frac{1}{n^2} \right)^3 \ \frac{max\left(l,l'\right)}{2l+1} \ |R_{n,l}^{n',l'}|^2.
\end{equation}
An extensive literature exists on how to calculate the radial integral, $R_{n,l}^{n',l'}$.  Gordon (1929) first provided exact expressions for it, in terms of hypergeometric functions.  Multiple authors have since developed different ways of computing this formula, mostly in tabular form (e.g., Green, Rush \& Chandler 1957;  Goldwire 1968; Menzel 1969; Khandelwal \& Fitchard 1972).  Malik, Malik \& Varma (1991) rewrote Gordon's formulae in terms of associated Laguerre polynomials, yielding
\begin{equation}
R_{n,l}^{n',l-1} = F_1 F_2 \sum_{\lambda=\lambda_0}^{n-l-1} \left({n+l}\atop\lambda\right) \left({n'+l-1}\atop{n'-n+\lambda+2}\right) \left({n+l-\lambda-3}\atop{2l-2}\right) u^\lambda (T_1+T_2+T_3+T_4+T_5)
\label{eq:radialminus}
\end{equation}
and
\begin{equation}
R_{n,l}^{n',l+1} = F_2 F_3 \sum_{\lambda=\lambda_0}^{n-l-1} \left({n+l}\atop\lambda\right) \left({n'+l+1}\atop{n'-n+\lambda+2}\right) \left({n+l-\lambda-1}\atop{2l}\right) u^\lambda (t_1+t_2+t_3+t_4+T_5).
\label{eq:radialplus}
\end{equation}
where $\lambda_0=max\left(0,n-n'-2\right)$ and $\left(a \atop b\right)=a!/(a-b)!~b!$ for any $a,b \in \mathbb{Z}$ is the binomial coefficient.

The various functions used in equations (\ref{eq:radialminus}) and (\ref{eq:radialplus}) are
\begin{eqnarray}
F_1 = \left[ \left({n+l}\atop{2l-2}\right) \left({n'+l-1}\atop{2l-2}\right) (n-l+2) (n-l+1) (n-l) (n'-l+1)\right]^{-1/2}, \nonumber\\
F_2 = 4^n \left[ \frac{nn'}{(n+n')^2} \right]^{n+1} \left(\frac{n'-n}{n'+n}\right)^{n'-n}, \nonumber\\
F_3 = \left[ \left({n'+l+1}\atop{2l}\right) \left({n+l}\atop{2l}\right) (n'-l+1) (n'-l) (n'-l-1) (n-1) \right]^{-1/2}, \nonumber\\
T_1 = -y^2 (j+l)(j+l-1)(j+l-2)(j-l-1), \nonumber\\
T_2 = 2y(j+l)(j+l-1)(2j-l-1)(n'-j+2), \nonumber\\
T_3 = -6j(j+l)(n'-j+1)(n'-j+2), \nonumber\\
T_4 = \frac{2}{y}(2j+l+1)(n'-j)(n'-j+1)(n'-j+2), \nonumber\\
T_5 = -\frac{1}{y^2}(n'-j-1)(n'-j)(n'-j+1)(n'-j+2), \nonumber\\
\end{eqnarray}
and
\begin{eqnarray}
t_1 = -y^2 (j+l)(j-l-3)(j-l-2)(j-l-1), \nonumber\\
t_2 = 2y(n'-j+2)(j-l-1)(j-l-2)(2j+l), \nonumber\\
t_3 = -6j(n'-j+1)(n'-j+2)(j-l-1), \nonumber\\
t_4 = \frac{2}{y}(n'-j)(n'-j+1)(n'-j+2)(2j-l), \nonumber\\
u=-\frac{(n-n')^2}{4nn'}, \nonumber\\
y=\frac{n'-n}{2n}, \nonumber\\
j=n-\lambda.
\end{eqnarray}
where we note that the last term in $F_1$ was erroneously written as $l$ in Malik, Malik \& Varma (1991).  With the correct version, equation (\ref{eq:radialminus}) reduces to the following, exact form (Hoang-Binh 1990):
\begin{equation}
R_{n,n-1}^{n-1,n-2} = \sqrt{(2n-1)(2n-2)}~ \frac{n(n-1)}{2n-1} \left[ \frac{4n(n-1)}{(2n-1)^2} \right]^n.
\end{equation}

\begin{figure}
\begin{center}
\includegraphics[width=6in]{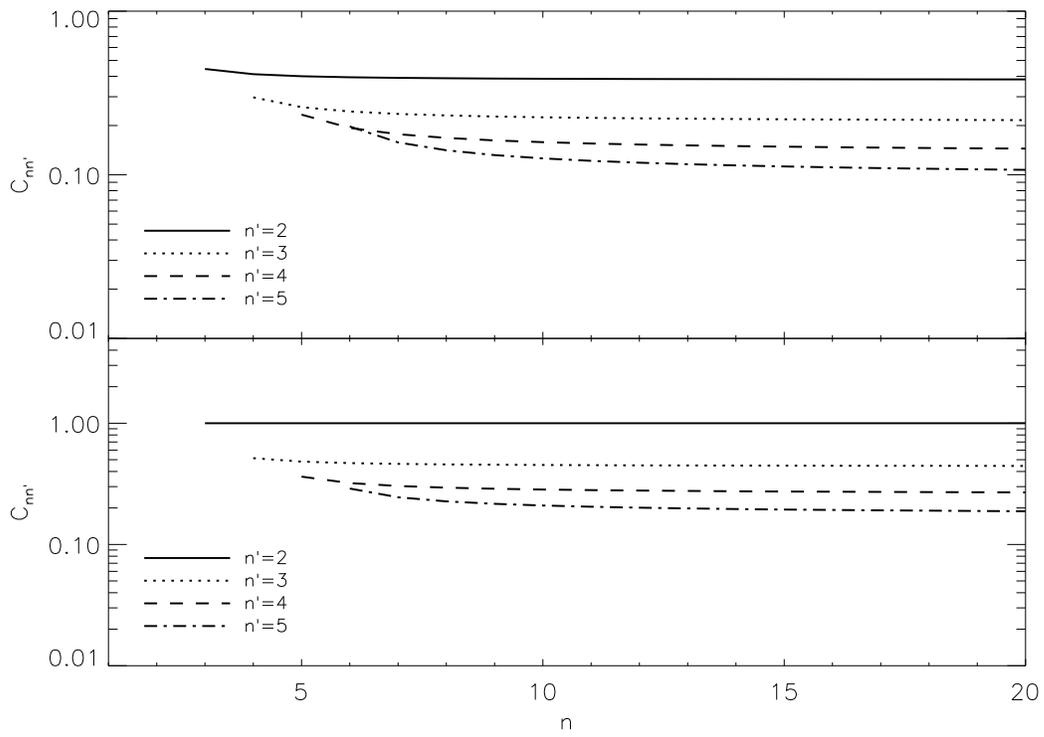}
\end{center}
\caption{Cascade matrices for Case A (top) and B (bottom).  Each matrix element, $C_{nn'}$, is the probability that the population of $n$ is followed by a transition to $n'<n$ via all possible cascade routes.}
\label{fig:cascade}
\end{figure}

Finally, we display the resulting cascade matrices in (Fig. \ref{fig:cascade}).  We have assumed that the excitations are statistically distributed among the angular momentum states.  We note that this is a reasonable assumption for charge transfer into excited states, but not necessarily so for impact excitation, which favors $np$ states, especially at high energies.


\begin{table}
\begin{center}
\caption{FWHM of Broad Component of H$\alpha$, $I_b/I_n$ and Predicted Shock Velocities}
\label{tab:fwhm}
\begin{tabular}{lcccccc}
\tableline\tableline
\multicolumn{1}{c}{SNR} & \multicolumn{1}{c}{Reference} & \multicolumn{1}{c}{FWHM} & \multicolumn{1}{c}{$v_s$ cited$^\dagger$} & \multicolumn{1}{c}{$I_b/I_n$ cited}  & \multicolumn{1}{c}{$v_s$ (Model N)$^\ddagger$} & \multicolumn{1}{c}{$v_s$ (Model F)$^\ddagger$}\\
\multicolumn{1}{c}{} & \multicolumn{1}{c}{} & \multicolumn{1}{c}{(km s$^{-1}$)} & \multicolumn{1}{c}{(km s$^{-1}$)}&  & \multicolumn{1}{c}{(km s$^{-1}$)} & \multicolumn{1}{c}{(km s$^{-1}$)}\\
\tableline
SN 1006 & G02 & 2290 $\pm$ 80 & 2865 --- 3580 & 0.84$^{+0.03}_{-0.01}$ & 2509 $\pm$ 111 & 2981 $\pm$ 133\\
SN 1006 & S91 & 2310 $\pm$ 210 & 2400 --- 3240 & 0.73 $\pm$ 0.06 & 2537 $\pm$ 292 & 3014 $\pm$ 354\\
SN 1006 &  KWC87 & 2600 $\pm$ 100 & 2800 --- 3870 & 0.77 $\pm$ 0.08  & 2940 $\pm$ 149 & 3545 $\pm$ 183\\
\tableline
Kepler & BLV91 & 1500 & 1550 --- 2000 & 0.72 $\pm$ 0.37 & 1528 & 1809\\
Kepler & F89 & 1750 $\pm$ 200 & 1670 --- 2800 & 1.1 $\pm$ 0.25 & 1806 $\pm$ 226 & 2154 $\pm$ 294\\
\tableline
Tycho & G01 & 1765 $\pm$ 110 & 1940 --- 3010 & 0.67 $\pm$ 0.1 & 1823 $\pm$ 122 & 2177 $\pm$ 169\\
Tycho & S91 & 1900 $\pm$ 300 & 1850 --- 2500 & 0.77 $\pm$ 0.09 & 1973 $\pm$ 372 & 2383 $\pm$ 453\\
Tycho & KWC87 & 1800 $\pm$ 100 & 1930 --- 2670 & 1.08 $\pm$ 0.16 & 1862 $\pm$ 111 & 2230 $\pm$ 153\\
\tableline
RCW 86 & G01 & 562 $\pm$ 18 & 545 --- 793 & 1.18 $\pm$ 0.03 & 536 $\pm$ 18 & 641 $\pm$ 21\\
Cygnus & G01 & 262 $\pm$ 32 & 235 --- 395 & 0.59 $\pm$ 0.3 & 247 $\pm$ 30 & 296 $\pm$ 36\\
0505---67.9 & S91 & 580 $\pm$ 70 & 480 --- 640 & $\gtrsim$ 0.7 & 554 $\pm$ 69 & 662 $\pm$ 82\\
0519---69.0 & S91 & 1300 $\pm$ 200 & 1180 --- 1580 &  0.8 $\pm$ 0.2 & 1305 $\pm$ 223 & 1554 $\pm$ 255\\
0548---70.4 & S91 & 760 $\pm$ 140 & 670 --- 890 & 1.1 $\pm$ 0.2 & 732 $\pm$ 140 & 874 $\pm$ 168\\
\tableline
\end{tabular}
\end{center}
$^\dagger$ Range of shock velocities quoted for zero and full electron-ion equilibration.\\
$^\ddagger$ Derived for edge-wise shock fronts.\\
\end{table}

\end{document}